\documentclass[12pt]{amsart}
\usepackage{amsthm,amsmath,amssymb,url}
\usepackage{xcolor}
\usepackage[top=15mm, bottom=20mm, left=30mm, right=30mm]{geometry}

\usepackage{hyperref}
\usepackage{fourier}

\usepackage{graphicx}
\usepackage{enumerate}

\theoremstyle{plain}

\newcommand{\mw}[1]{\textcolor{red}{\textit{#1}}}

\begin{document} 
\title{Multi-district preference modelling} 
\author{Geoffrey Pritchard} 
\address{University of Auckland, Auckland, NEW ZEALAND} 
\email{g.pritchard@auckland.ac.nz}
\author{Mark C. Wilson} 
\address{University of Auckland, Auckland, NEW ZEALAND} 
\email{mcw@cs.auckland.ac.nz} 
\thanks{The authors acknowledge the support of the New Zealand Marsden Fund. The authors thank B. Grofman for useful discussions.}

\begin{abstract}
Generating realistic artificial preference distributions is an important part of any simulation analysis of electoral systems. While this has been discussed in some detail in the context of  a single electoral district, many electoral systems of interest are based on multiple districts. Neither treating preferences between districts as independent nor ignoring the district structure yields satisfactory results. We present a model based on an extension of the classic Eggenberger-P\'{o}lya urn, in which each district is represented by an urn and there is correlation between urns. We show in detail that this procedure has a small number of tunable parameters, is computationally efficient, and produces ``realistic-looking" 
distributions. We intend to use it in further studies of electoral systems.

\end{abstract}


\maketitle

\section{Introduction} 
In order to test the average-case performance of electoral systems, or to test models of electoral phenomena such as inter-election swing, it is useful to be able to generate artificial preference distributions for a society. 

We are particularly interested in multi-winner elections of parliamentary type, in which the electorate is partitioned into districts, one or more representatives are chosen from each district, and these representatives together make up the parliament. It is important neither to assume  independence of districts nor to ignore the districts altogether, since in each case we obtain very unrealistic results. For example, if simulating plurality voting in single-winner districts (``First Past the Post") where party A has nationally 55\% of the vote, the latter assumption will typically give 100\% of seats to that party, while the former will be close to a proportional representation outcome in which it wins 55\% of seats.

In social choice theory (basically corresponding to the single-district case), the theoretical performance of voting systems, for example with respect to manipulability, has been studied in detail using various preference distributions \cite{BeLe1994, RGMT2006}. This kind of work may be useful for studying, for example, party list proportional systems. However, we are not aware of any work for district-based electoral systems.
In this article we present a general method for generating artificial preference data in districts which 
has a small number of tunable parameters, is computationally efficient, and produces ``realistic-looking" 
distributions. We cover only the case of plurality ballots.

\section{Basic models}
\label{s:models}

\subsection{Eggenberger-P\'{o}lya urn models}
\label{ss:PE}

We review the basic abstract framework, which in its essentials goes back to Eggenberger and P\'olya in \cite{EP23}, although they in turn drew on the work of still earlier authors. Let $K$ be a nonnegative integer. Consider a container (\emph{urn}) into which we place initially some number $a_i$ of indistinguishable \emph{balls} of each colour $i$, such that the total number of balls is finite. At each step we randomly choose a ball from the urn, and replace it together with $K$ balls of the same colour.  When $K=0$, this is simply sampling with replacement, but for larger values of $K$, the number of balls increases at each iteration with a bias toward colours that already occur frequently.

This process has been used to model contagion. In the context of preferences, it has been widely used \cite{BeLe1994}.
If we assign a different colour to each possible agent preference, the urn model can be interpreted as describing what happens when social influence occurs. In particular when $s=1$, we can interpret it as adding new agents one at a time, and each agent imitating the preference of an existing agent with probability proportional to the frequency of that agent's preference in the population. 

The initial distribution of balls of course influences the distribution at later times. Most obviously, no colour not already present in the initial distribution can ever be present in the urn.

\subsection{The urn model with multiple districts}
\label{ss:PEdist}

We extend the basic model above to one with multiple urns. Let $p$ be a probability, that is, a real number between $0$ and $1$ (inclusive). We have $N$ urns, one for each district. In each urn there is an initial distribution of balls as above --- any possible overall distribution between districts and within districts is possible. At each step, we first choose an urn $u$ uniformly at random.
With probability $p$, we choose to draw a ball from some other urn (chosen uniformly at random
from among all urns other than $u$); otherwise we draw a ball from $u$. When the ball has been
drawn, it is returned to the urn from whence it came, and an additional ball of the same colour
is added to $u$.

\if01
Figure~\ref{fig:eg} shows a representative example where $N=10, p = 0.3, m = 2$ and the initial conditions are that Party 1 has 2 voters in each of the first 5 districts and 1 in each of the others, while 
Party 2 has 1 voter in each of the first 5 districts and 2 in each of the others.
\fi

\section{Theoretical properties of the model}
\label{s:theory}

The known theoretical properties of the single Eggenberger-P\'{o}lya urn
can provide some insight into the behaviour of more complex multiple-urn models.
Suppose a single urn initially contains $a_1,\ldots,a_m$ balls of
colours $1,\ldots,m$ respectively, and each ball drawn is replaced along with one
more ball of the same colour.
Let $X_1,X_2,\ldots$ be random variables giving the colours of the balls drawn, and
$Y_1,Y_2,\ldots$ be random $m$-vectors giving the proportions of the colours after
each draw ({\it i.e.} $Y_n(k)$ is the proportion of the balls that have colour $k$
after the first $n$ draws).
Then
\begin{enumerate}
\item With probability 1, the sequence $(Y_n)$ converges to a limit $Y$;
\item $Y$ is a random variable with the Dirichlet$(a_1,\ldots,a_m)$ distribution;
\item Conditional on $Y$, the random variables $X_1,X_2,\ldots$ are independent
with probability distribution $Y$.
\end{enumerate}
(See \cite{BM73}.)
If there are only two colours, we may consider $Y_n$ to be scalar-valued
(since then $Y_n(2)$ is simply $1 - Y_n(1)$, adding no further information);
in this case the limit $Y$ has a Beta$(a_1,a_2)$ distribution.

A particularly simple special case occurs when $a_1=\cdots=a_m=1$ {\it i.e.} the
urn begins with one ball of each colour. The limiting Dirichlet$(1,\ldots,1)$
distribution is then the uniform distribution on the $m$-simplex;
for the two-colour case this is the uniform distribution on the interval $[0,1]$.
 
Turning to multiple urns, we can make some observations about what to expect.
Firstly, since each step adds a ball to one randomly selected urn, the urns will likely
end up containing different numbers of balls. However, once the number $n$ of steps taken
has become large, there will be similar numbers of balls in each urn (the differences
between the numbers of balls in different urns will be on the order of $\sqrt{n}$).
That is, the political districts represented by the urns will have similar numbers of voters.

Secondly, when $p=0$, the urns do not interact. Each urn behaves as a single Eggenberger-P\'{o}lya urn,
independently of the others.

Thirdly, when $p=1$ and $N$ is not small, the behaviour will be close to that exhibited in the case where there is a single large district, leading to substantial correlation between the ball counts in districts, for each colour.

\section{Evaluation}
\label{s:eval}

In order to test the realism of the model, we carried out extensive simulation using the model (C++ program available at \mw{give URL according to publisher instructions}). The baseline computations involved $N=100$, and we also performed some for $N=1000$. We concentrated on values of $m$ from $2$ to $4$, and varied $p$ from $0$ to $1$ in steps of $0.1$. We experimented with various initial conditions. For each case (value of $N,m,p$ and initial conditions) we ran $1000$ simulations.

We display the results using several types of graphs. First we fix a single party $A$ and show a histogram of the number of seats $S$ that it wins under FPP, and a histogram of its total (popular) vote share across districts. Next, for party $A$ we fix a district and show a histogram of the vote share in that district. We also compute the correlation between $A$'s share of the total vote in the first 50 (arbitrarily labelled ``north") districts and in the second 50 (labelled ``south") districts. 

\subsection{Two parties}
\label{ss:two}

We fix $m=2$. 

\subsubsection{Symmetric initial conditions}
\label{sss:symm}

\begin{figure*}
\caption{District vote share distribution when $p=0$ (L) and $p=0.2$ (R)}
\label{fig:votes_init11}
\includegraphics[width=7cm]{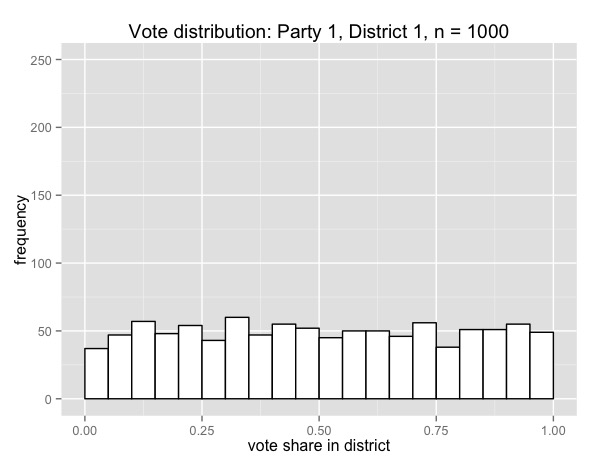}
\includegraphics[width=7cm]{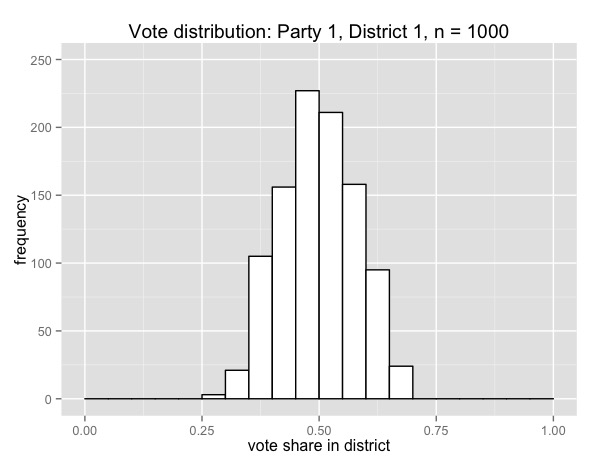}
\end{figure*}

\begin{figure*}
\caption{Seat share distribution when $p=0$ (L) and $p=0.2$ (R)}
\label{fig:seats_init11}
\includegraphics[width=7cm]{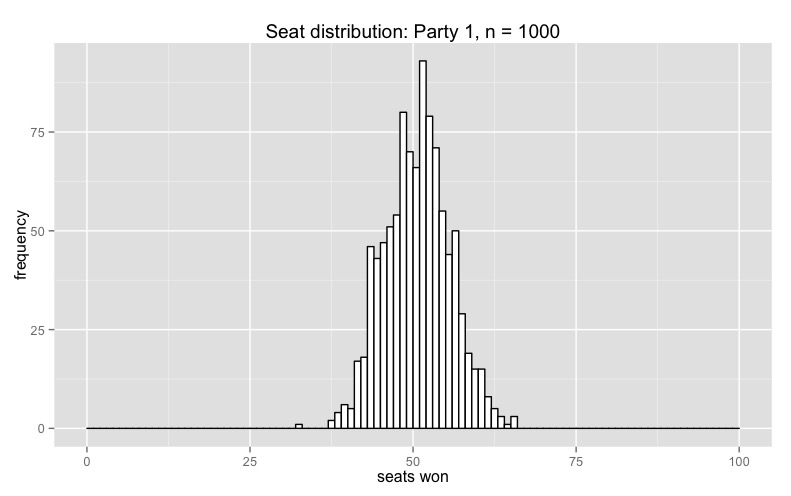}
\includegraphics[width=7cm]{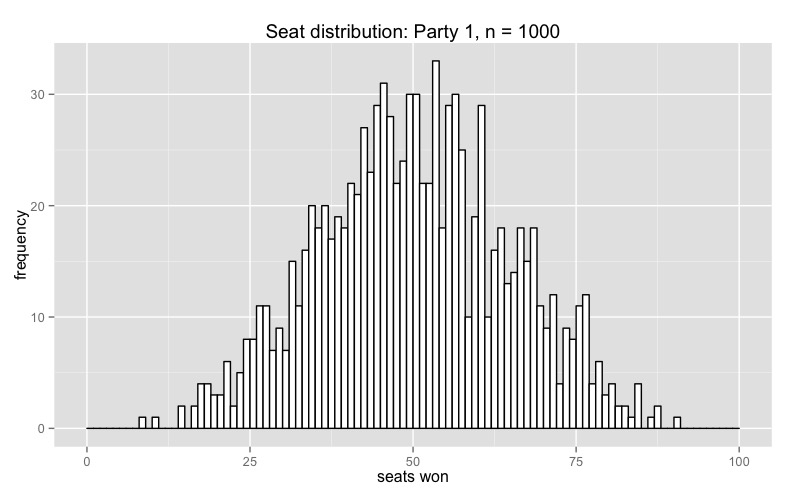}
\end{figure*}

\begin{figure*}
\caption{North-south correlation  when $p=0$ (L) and $p=0.2$ (R)}
\label{fig:northsouth_init11}
\includegraphics[width=7cm]{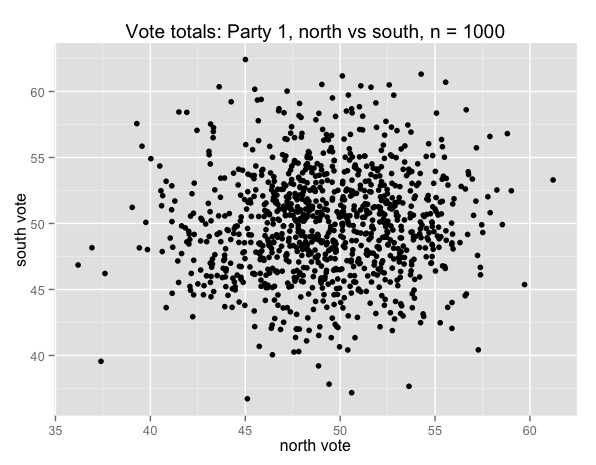}
\includegraphics[width=7cm]{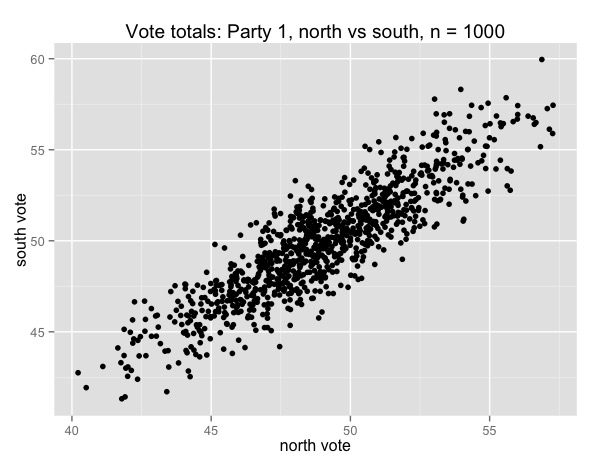}
\end{figure*}

\begin{figure*}
\caption{Vote - seat mapping when $p=0$ (L) and $p=0.2$ (R)}
\label{fig:seatvote_init11}
\includegraphics[width=7cm]{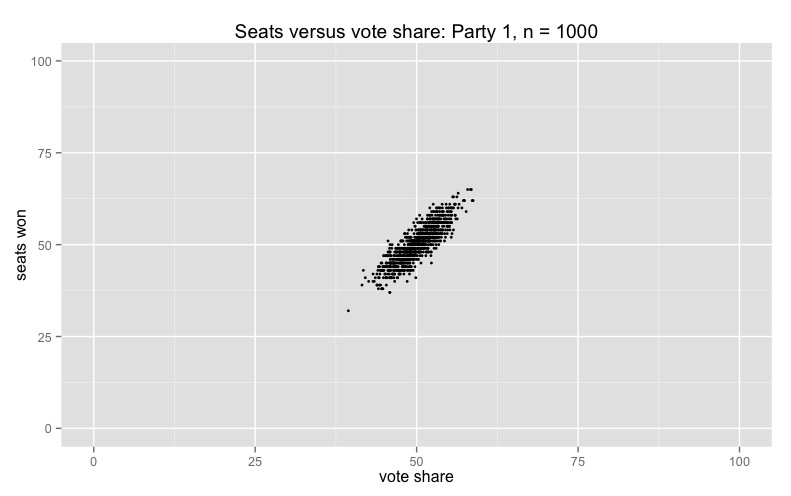}
\includegraphics[width=7cm]{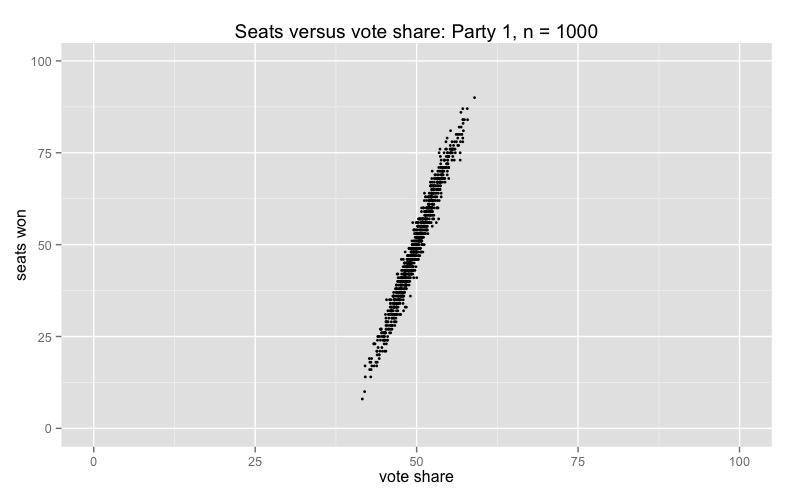}
\end{figure*}

Here we started with one voter of each party in each district. When $p=0$, the distribution of the vote share of Party 1 in District 1 is shown in the left picture in Figure~\ref{fig:votes_init11} (this is a sample from a uniform distribution) and the distribution of this party's seat share in the left picture of Figure~\ref{fig:seats_init11} (this is a sample from a binomial distribution). As expected, we see no correlation between the party's vote in the northern and southern districts --- see the left picture in Figure~\ref{fig:northsouth_init11}. The left picture in Figure~\ref{fig:seatvote_init11} shows a positive association between votes and seats, again as expected. 

\begin{figure*}
\caption{Popular vote share distribution when $p=0$ (L) and $p=0.2$ (R)}
\label{fig:popvote_init11}
\includegraphics[width=7cm]{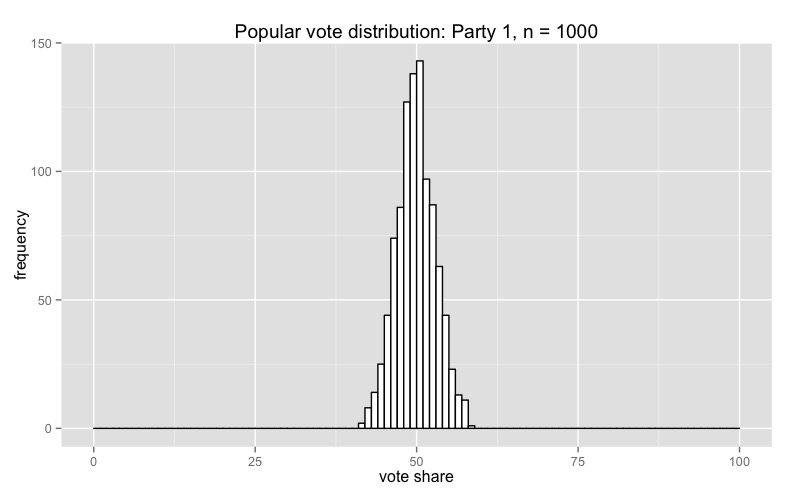}
\includegraphics[width=7cm]{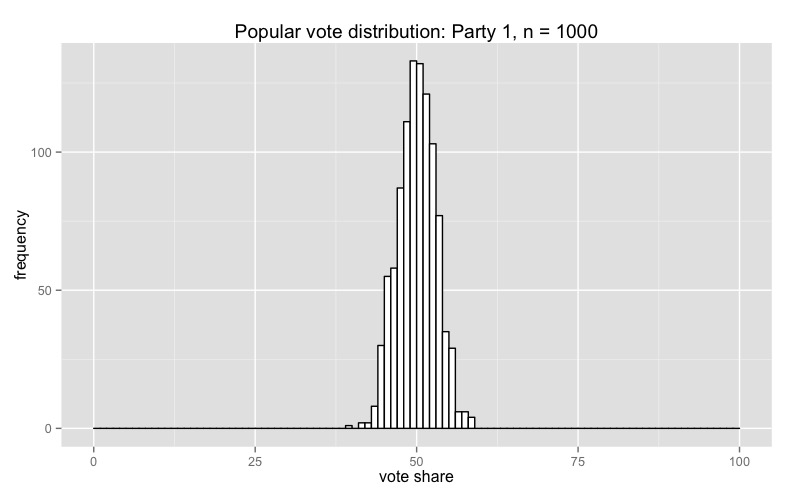}
\end{figure*}

Increasing the value of $p$ to $0.2$ leads to quite different results
(see the right-hand pictures in Figures \ref{fig:votes_init11}-\ref{fig:seatvote_init11}).
The distribution of the party's vote in a given district becomes more concentrated; the influence of other districts now makes it much less likely that this district will deliver a result wildly different from the national average.
At the same time, the seat distribution becomes \emph{less} concentrated. The stabilization of the within-district vote puts many more districts in play, while inter-district imitation has a self-reinforcing effect: a high vote in one district not only wins that seat, but improves the party's prospects in every other district as well. Another view of this effect is provided by the north-south correlation graph (Figure \ref{fig:northsouth_init11}). Also, the mapping of votes to seats has a larger slope, a point we shall return to in Section \ref{ss:cube}.

\begin{figure*}
\caption{The case $p=1$: seats (L) and seats vs votes (R)}
\label{fig:seatvote_init11_10}
\includegraphics[width=7cm]{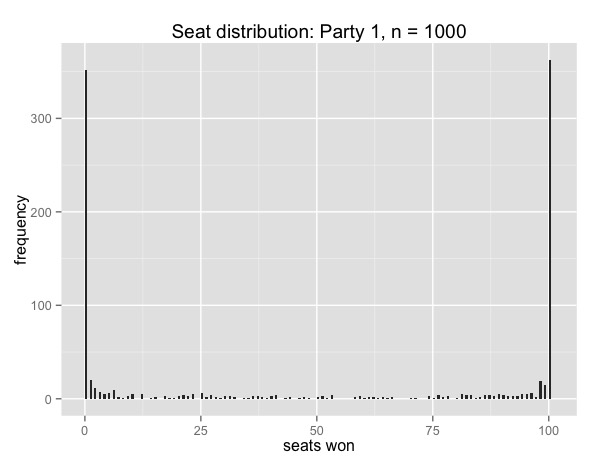}
\includegraphics[width=7cm]{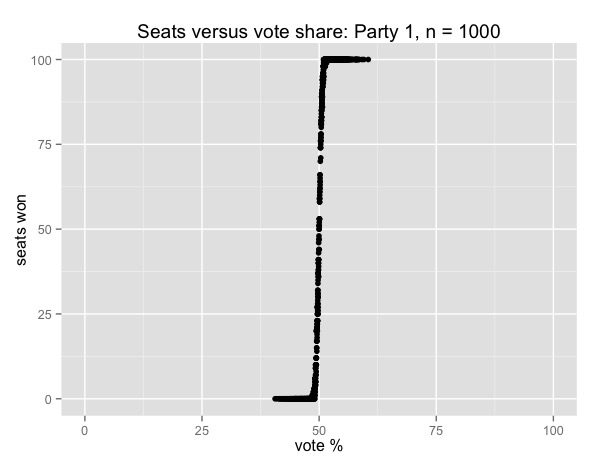}
\end{figure*}

As $p$ increases, the trend observed above continues. Figure~\ref{fig:seatvote_init11_10} shows the behaviour in the extreme case $p=1$: the entire electorate is now behaving (approximately) like a single well-mixed district, and it is quite likely that all of the parliamentary seats will be won by the same party. (Due to the symmetry of the situation, each of our two parties has the same probability of achieving such a total victory.) Note that the vote shares (both nationally and in each district) are still quite close to 50-50; it is the lack of diversity among districts that makes such lopsided seat-shares possible. 
This case could be thought of as resembling a single P\'{o}lya urn with an initial content of 50 balls of each colour.

\begin{figure*}
\caption{Initial conditions $2:2$. District vote (L), popular vote (M) and seat (R) distribution when $p=0$.}
\label{fig:votes_22}
\includegraphics[width=5cm]{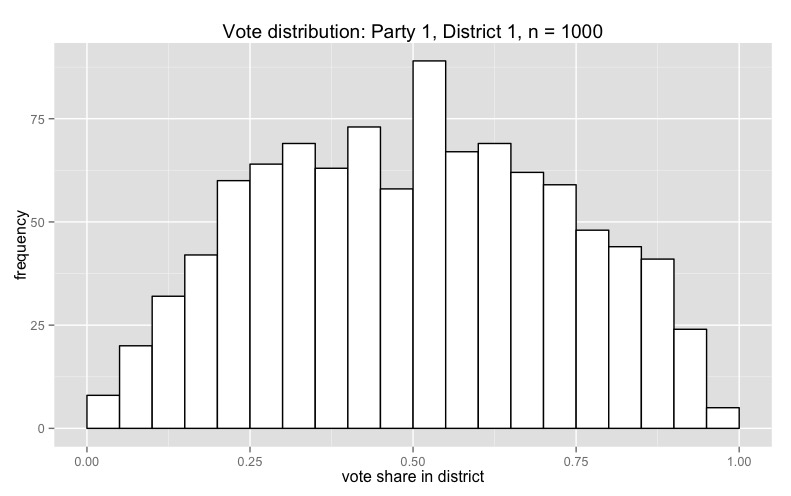}
\includegraphics[width=5cm]{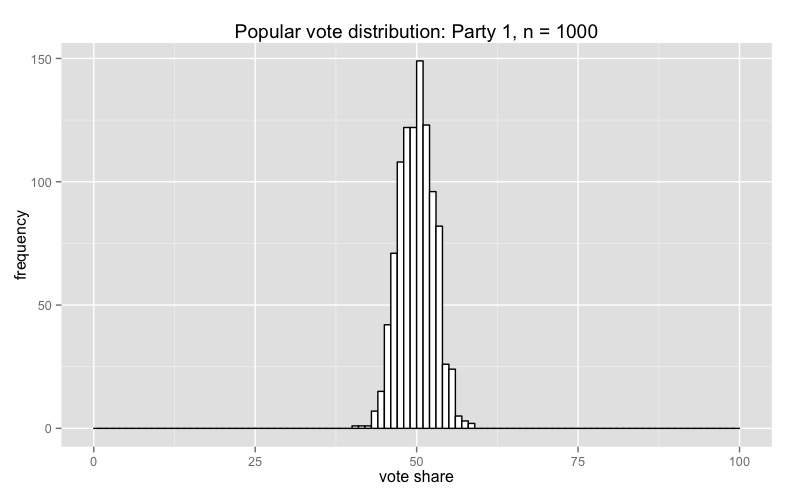}
\includegraphics[width=5cm]{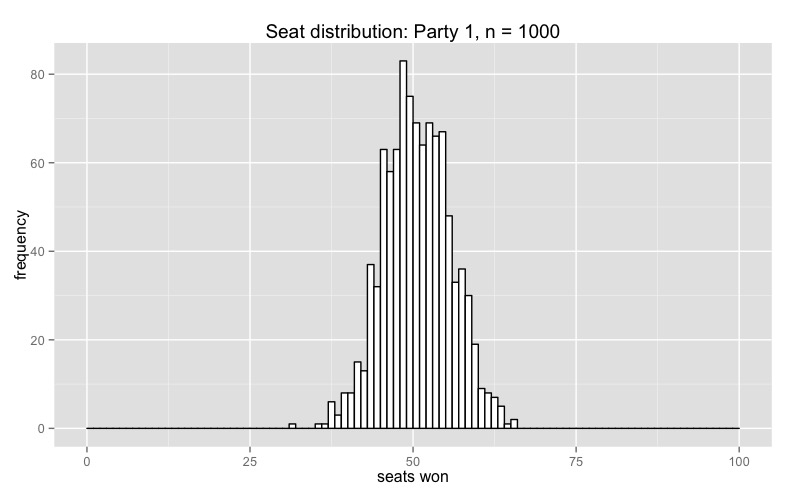}
\end{figure*}

\begin{figure*}
\caption{District vote (L), popular vote (M) and seat (R) distribution when $p=0.2$: initial conditions  $2:2$}
\label{fig:votes_22_02}
\includegraphics[width=5cm]{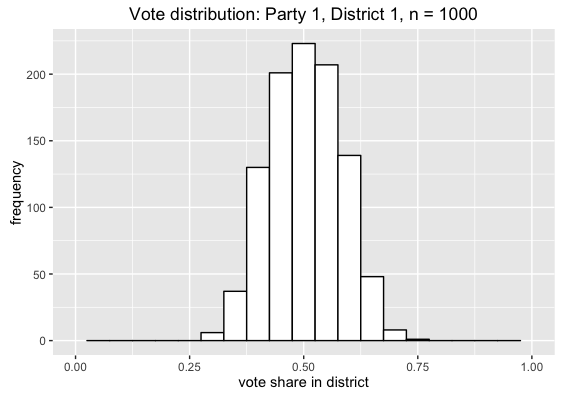}
\includegraphics[width=5cm]{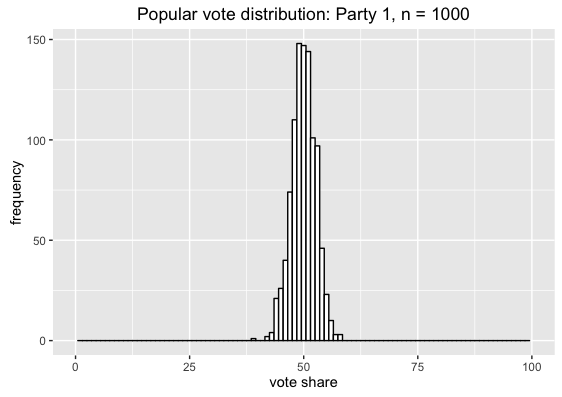}
\includegraphics[width=5cm]{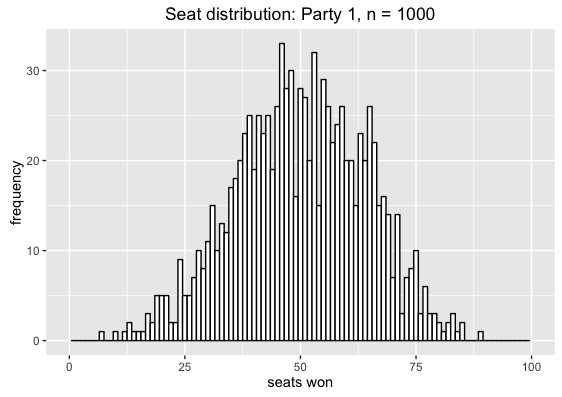}
\end{figure*}

\begin{figure*}
\caption{Vote - seat mapping when $p=0$ (L) and $p=0.2$ (R), initial conditions 2:2}
\label{fig:seatvote_init22}
\includegraphics[width=7cm]{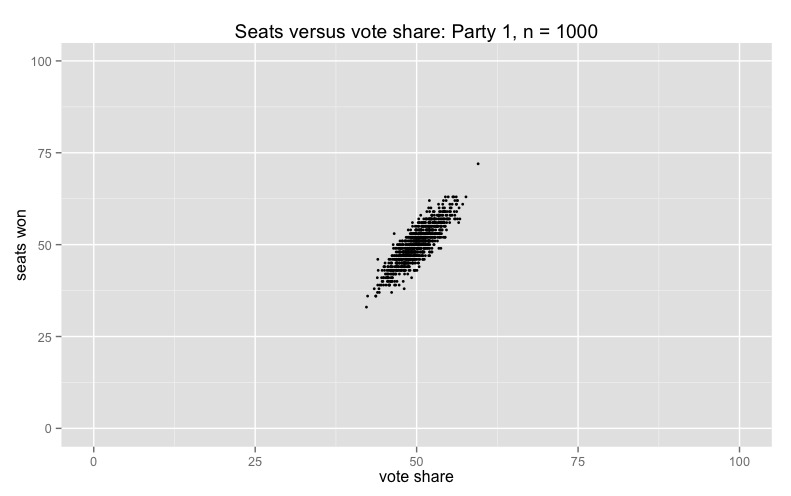}
\includegraphics[width=7cm]{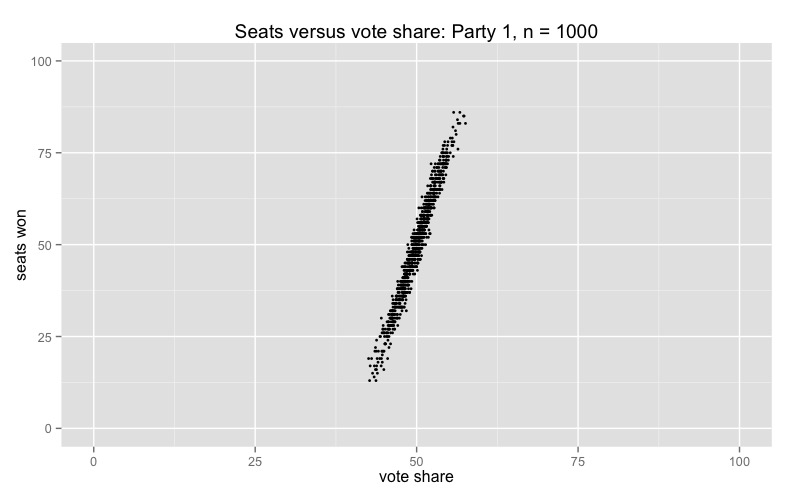}
\end{figure*}

We repeated the above experiments for the case where each party initially has 2 voters in each district (Figures~\ref{fig:votes_22} and \ref{fig:votes_22_02}). When $p=0$, the district vote distribution is Beta$(2,2)$ (density proportional to $x(1-x)$); consequently, the popular vote is a little less variable than in Figure~\ref{fig:popvote_init11}. The seat-share distribution is still binomial, unchanged from Figure~\ref{fig:seats_init11}, and the north-south correlation is still zero (graph omitted). The seat-vote relationship is shown in Figure~\ref{fig:seatvote_init22}. Introducing inter-district mimicry again has the effect of homogenizing districts, thereby making the seat-share distribution more variable. When we increase the initial number of voters
even more, the trend continues, with the district vote share and popular vote share both approaching a point mass at $1/2$.

\if01
\begin{figure}
\caption{District vote share distribution when $p=0$ (L) and $p=0.2$ (R), initial conditions 2:2}
\label{fig:votes_init22}
\includegraphics[width=7cm]{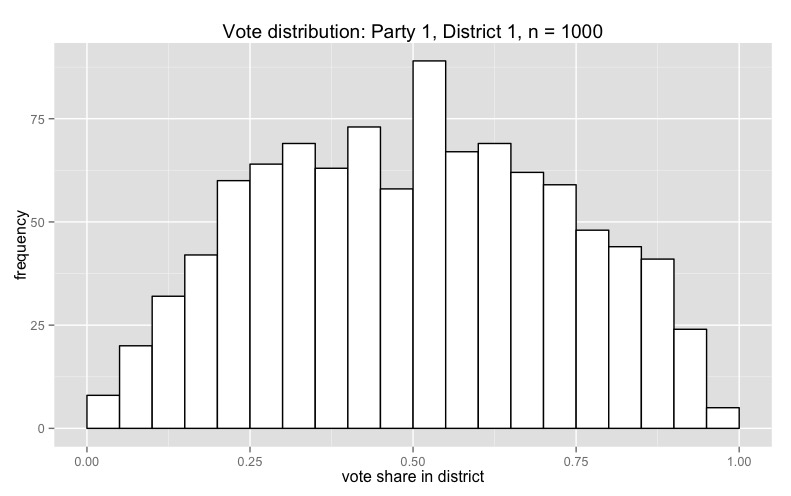}
\includegraphics[width=7cm]{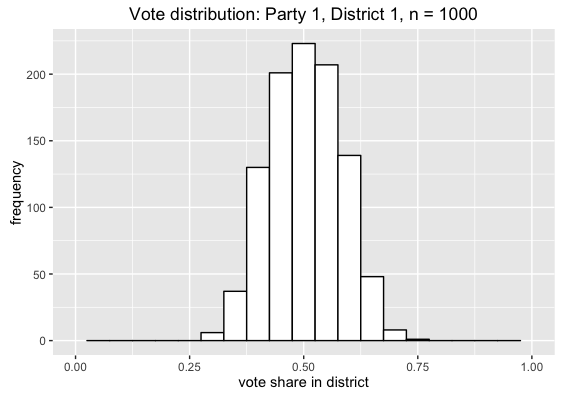}
\end{figure}

\begin{figure}
\caption{Popular vote share distribution when $p=0$ (L) and $p=0.2$ (R), initial conditions 2:2}
\label{fig:popvote_init22}
\includegraphics[width=7cm]{pictures/popvote_init22.jpeg}
\includegraphics[width=7cm]{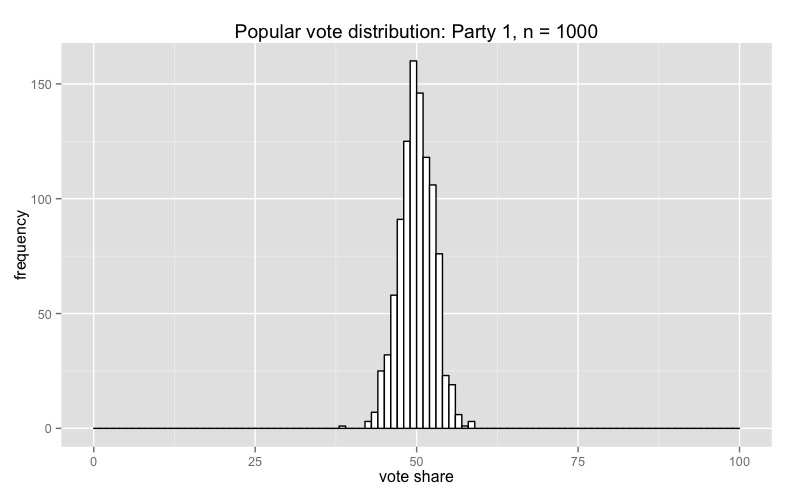}
\end{figure}

\begin{figure}
\caption{Seat share distribution when $p=0$ (L) and $p=0.2$ (R), initial conditions 2:2}
\label{fig:seats_init22}
\includegraphics[width=7cm]{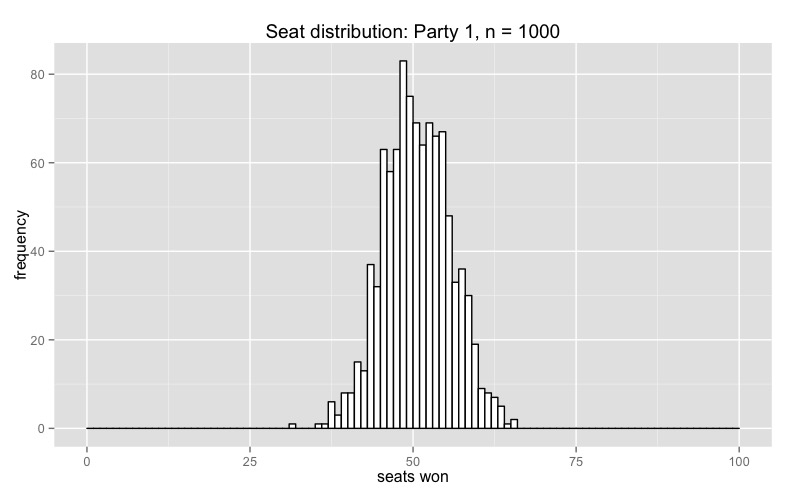}
\includegraphics[width=7cm]{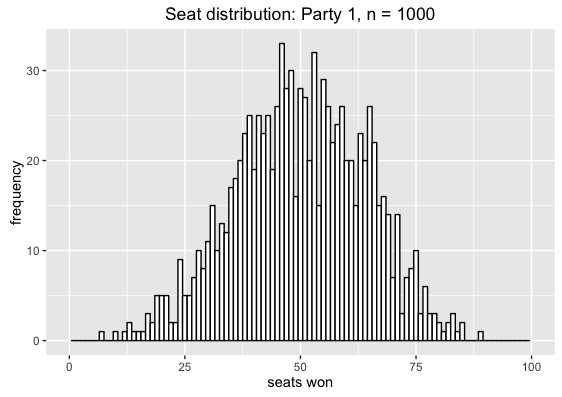}
\end{figure}

\fi

\begin{figure*}
\caption{District vote (L), popular vote (M) and seat (R) distribution when $p=0$: initial conditions  $1:1$, $K=5$}
\label{fig:votes_11K5}
\includegraphics[width=5cm]{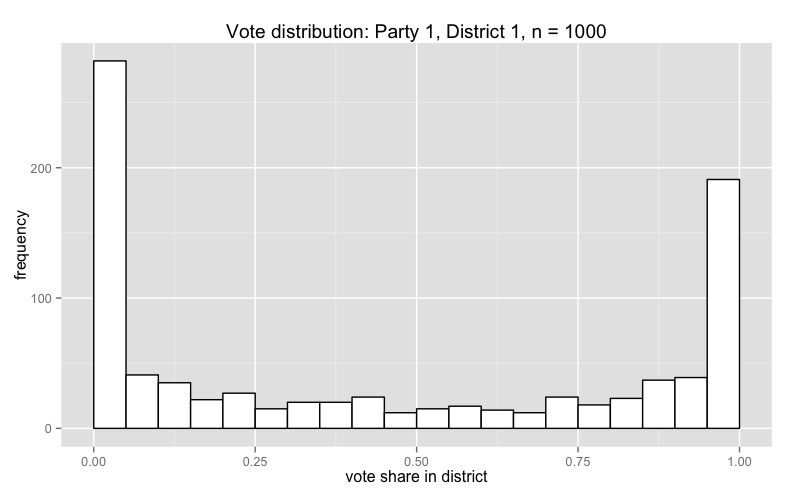}
\includegraphics[width=5cm]{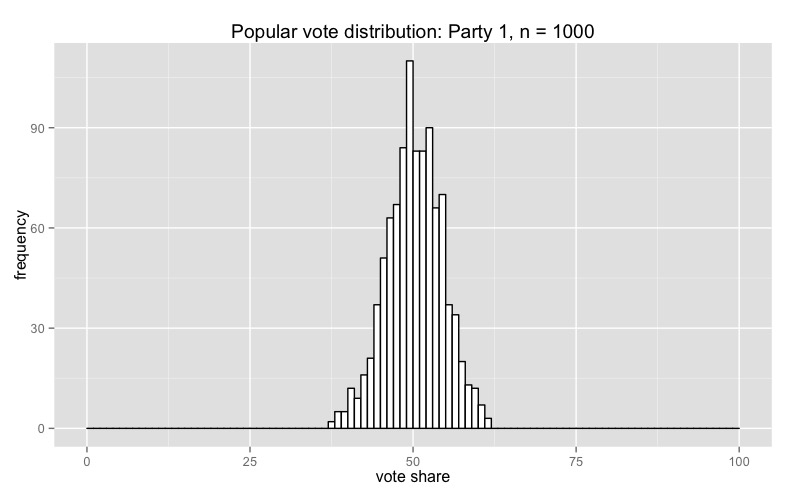}
\includegraphics[width=5cm]{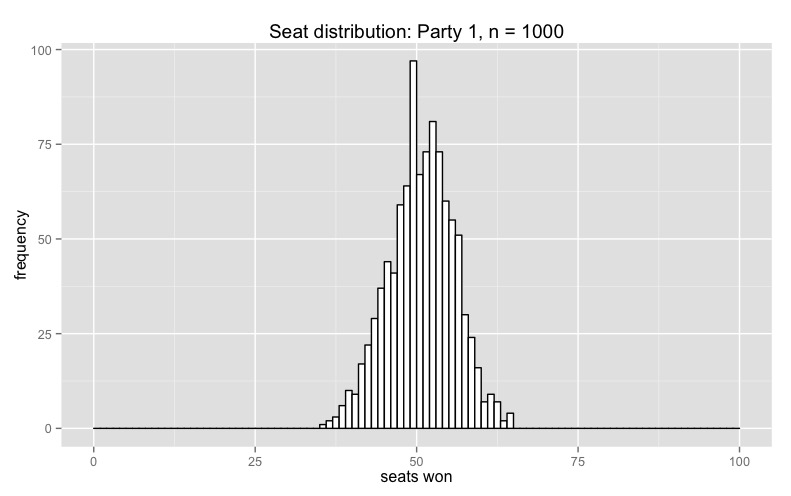}
\end{figure*}

\begin{figure*}
\caption{District vote (L), popular vote (M) and seat (R) distribution when $p=0.2$: initial conditions  $1:1$, $K=5$}
\label{fig:votes_11K5_02}
\includegraphics[width=5cm]{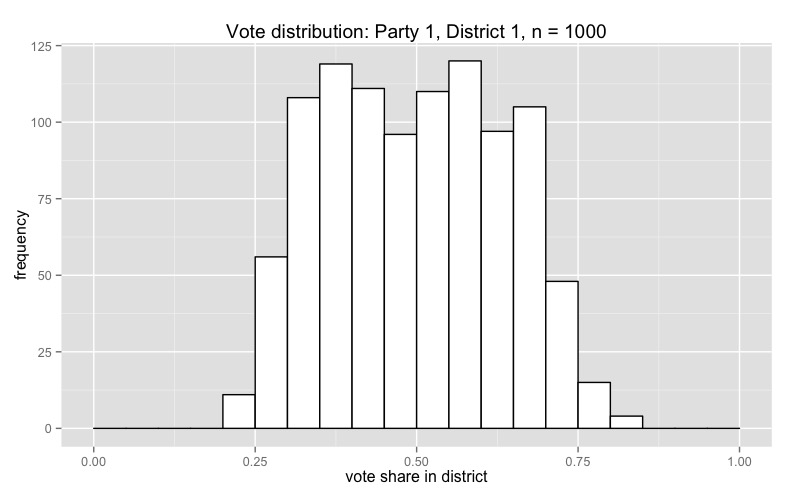}
\includegraphics[width=5cm]{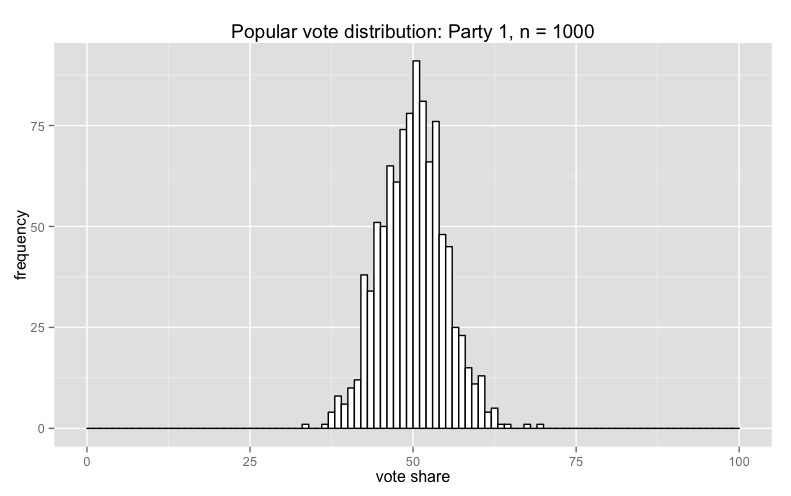}
\includegraphics[width=5cm]{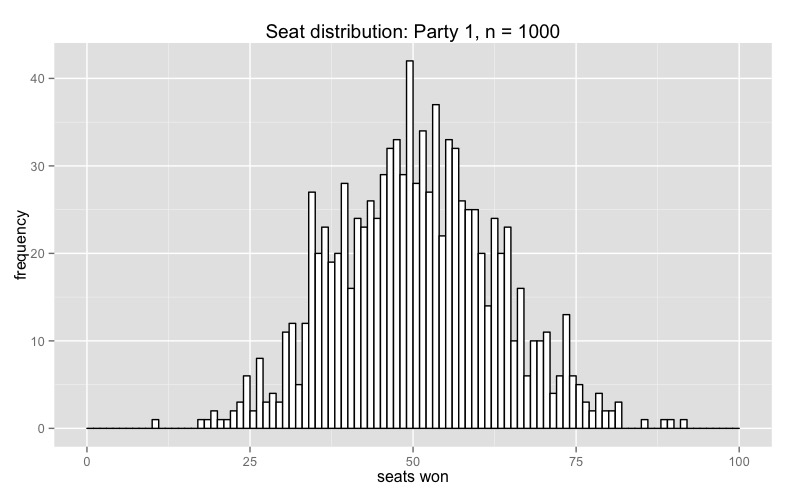}
\end{figure*}

\begin{figure*}
\caption{Vote - seat mapping when $p=0$ (L) and $p=0.2$ (R), initial conditions $1:1$, $K=5$}
\label{fig:seatvote_init11K5}
\includegraphics[width=7cm]{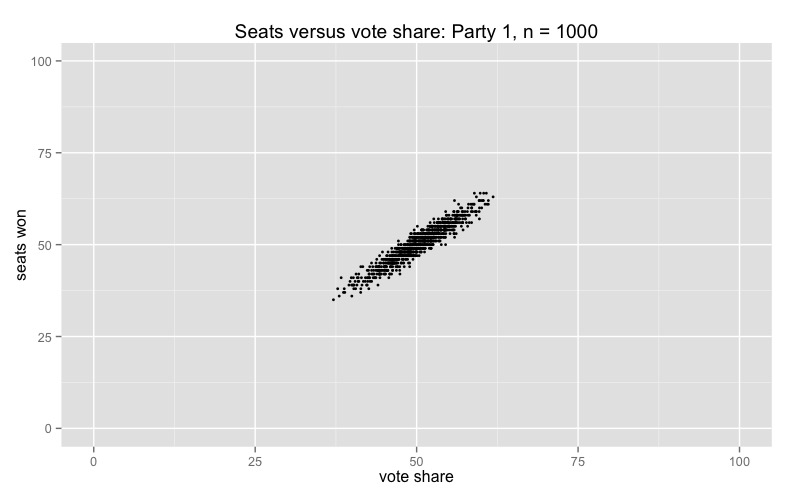}
\includegraphics[width=7cm]{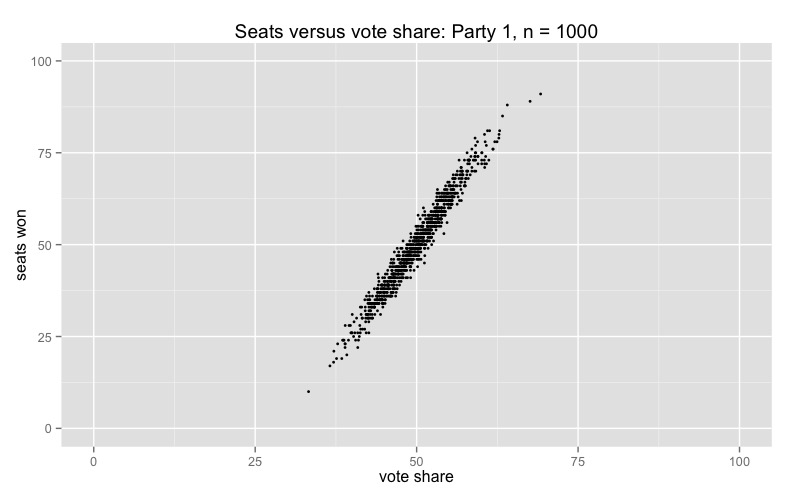}
\end{figure*}

Going in the other direction, we considered the case where the initial conditions can be rapidly forgotten as districts become swamped by imitated voters (alternatively, the initial number of voters is a small fraction of $1$). For example, we recomputed the above results when each party has 1 voter in each district initially, but $K=5$ new voters are added at each step. These are shown in Figures~\ref{fig:votes_11K5}--\ref{fig:seatvote_init11K5}. When $p=0$, note the very different vote share distribution  but the similarity of the other quantities presented. Also note that when $p=0.2$ we get results quite similar to the other cases presented above --- imitation washes out the effect of initial conditions.

\subsection{Other initial conditions}
\label{ss:asymm}

\begin{figure*}
\caption{District vote (L), popular vote (M) and seat (R) distribution when $p=0$: initial conditions  $2:1/1:2$}
\label{fig:votes_polarised}
\includegraphics[width=5cm]{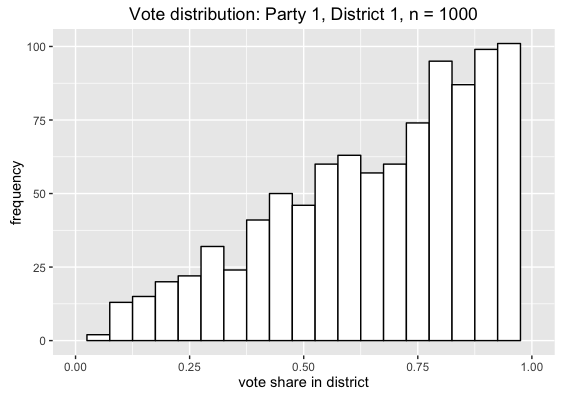}
\includegraphics[width=5cm]{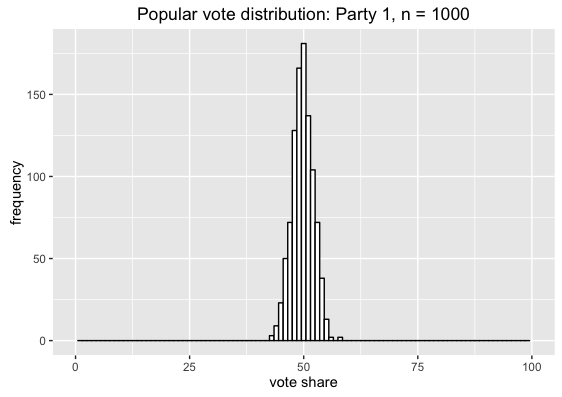}
\includegraphics[width=5cm]{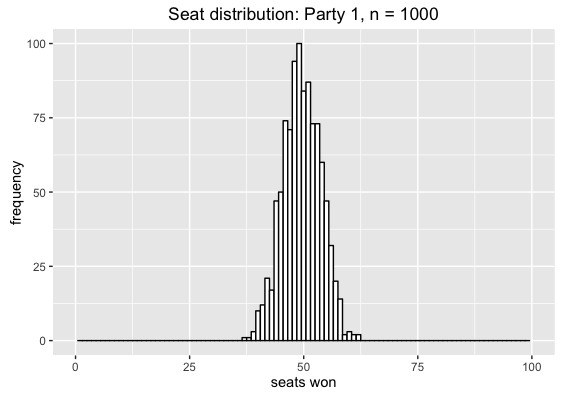}
\end{figure*}

We next considered initial conditions in which for the northern districts, Party 1 starts with the advantage of two initial voters to Party 2's single voter, while this advantage is reversed in the southern districts. In the absence of inter-district imitation (\emph{i.e.} $p=0$), this means that the district-level vote distribution is now Beta$(2,1)$, or triangular (Figure~\ref{fig:votes_polarised}, left panel), with each district having probability $\frac{3}{4}$ of being won by its dominant party. The resulting seat-share distribution resembles Figure~\ref{fig:seats_init11} (left panel), but with less variance. 

The polarisation of this electorate along geographical lines can be reduced by increasing the value of $p$. Imitation across the north-south boundary has the effect of making at least some districts more competitive, and increasing the variance of the seat-share distribution. This is perhaps a step towards a more realistic version of our model: real-world elections typically have many districts in which the mild initial dominance of one party can be overcome by national-level effects.

\begin{figure*}
\caption{District vote (L), popular vote (M) and seat (R) distribution when $p=0.2$: initial conditions  $2:1/1:2$}
\label{fig:votes_polarised_02}
\includegraphics[width=5cm]{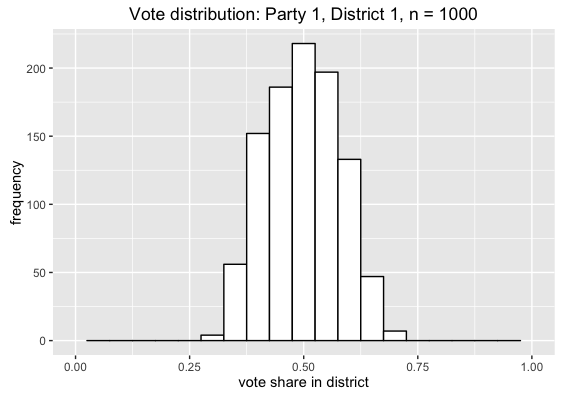}
\includegraphics[width=5cm]{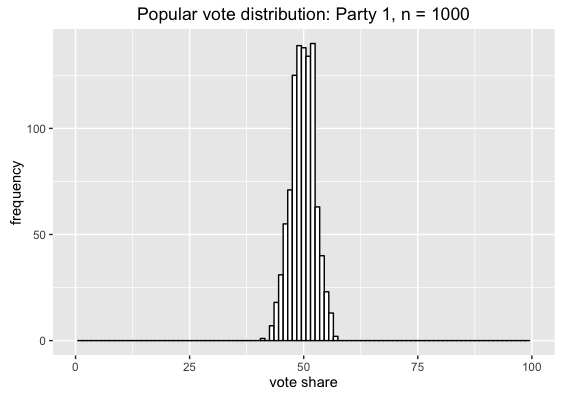}
\includegraphics[width=5cm]{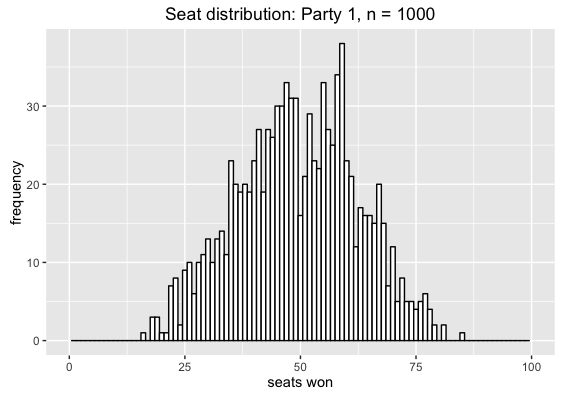}
\end{figure*}

We repeated the analysis with initial bias of $3:1$ instead of $2:1$, and the results were as expected. For $p=0$, there is a continuation of the trends in district vote and seat distribution, and a shift in the mean of the popular vote distribution but little change in spread. The slope of the seat-vote curve decreases. With $p=0.2$, the effect of the initial conditions has been largely washed out.

In summary, we have found that increasing $p$ has the effect of concentrating the vote distribution in a single district, leaving the popular vote distribution essential unchanged, and spreading out the seat distribution. Furthermore it increases north-south correlation and steepens the seat-vote curve. Changing the initial conditions of the process (but keeping the symmetry between parties overall) has a strong effect on the shape of the vote distribution in a given district, but does not change anything else substantially. 

\subsection{Modelling third parties}
\label{ss:regional}

In order to investigate scenarios involving a smaller third party competing with two roughly equal larger ones, we ran our model with the following sets of initial conditions.

\begin{enumerate}[(i)]
\item in 80 districts Parties 1,2,3 have respectively 0,2,2 voters; in 10 districts they have 1,2,2 and in 10 districts they have 2,1,1. 
\item as above, but in the last 10 districts (the ``regional base") the initial conditions are 3,1,1.
\item in all districts parties 1,2,3 have respectively 1,2,2 voters.
\end{enumerate}

\begin{figure*}
\caption{Seat shares for small party under three scenarios ($p=0.1$)}
\label{fig:3party}
\includegraphics[width=5cm]{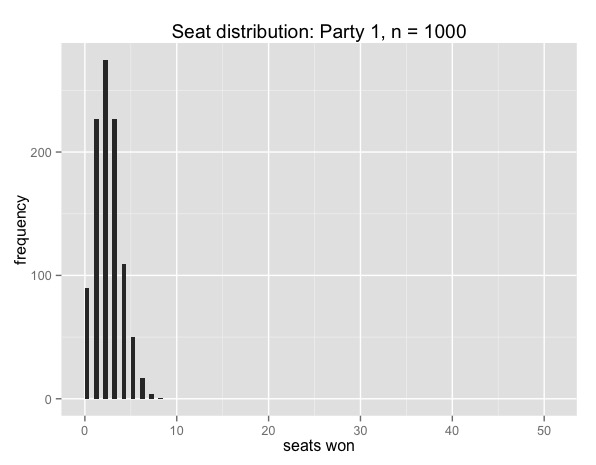}
\includegraphics[width=5cm]{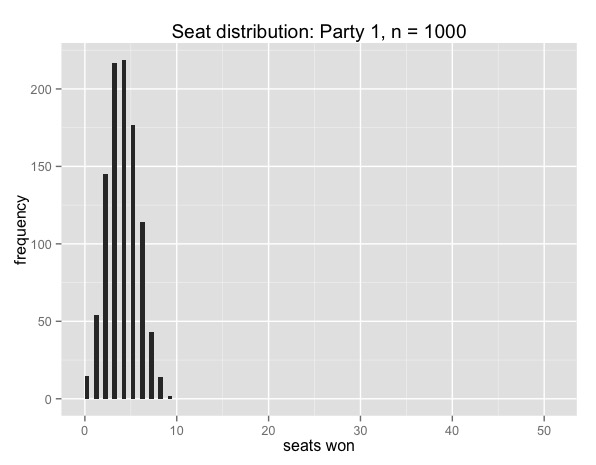}
\includegraphics[width=5cm]{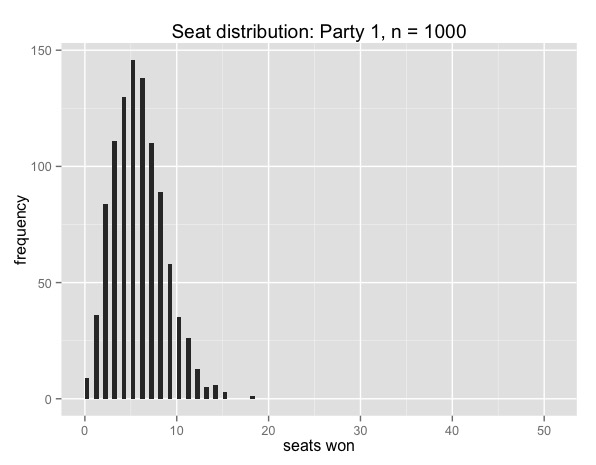}
\end{figure*}

We expect that the small party will fail to become large. When $p=0$, it cannot extend to any of the first 80 seats in the first two cases. As $p$ increases, the imitation means that although it attracts voters in districts away from its base, it loses more in its base to the other parties.
We present graphs on seat distribution for Party 1 under the three scenarios in Figure~\ref{fig:3party}.

\section{Applications}
\label{s:app}

\subsection{The cube law}
\label{ss:cube}

It has long been observed (see {\it e.g.} \cite{KS50}) that in two-party electoral systems where each district elects a single representative by simple plurality
voting (``First Past the Post", or FPP), there is a relationship between the fraction $x$ of all the votes and the fraction $y$ of all the districts won by a party along the lines of
\begin{equation}
 \frac{y}{1-y} \;\approx\; \left(\frac{x}{1-x}\right)^k .
 \label{eq:cubelaw}
\end{equation}
When $k=3$ this is sometimes known as the ``cube law'', or ``law of the cubic proportion''. For a more comprehensive discussion of this ``law'' and some alternatives to it, see \cite{T73, Taag1973, Taag1986, GT79}.

Taking (\ref{eq:cubelaw}) as an equality gives
\begin{equation}
\label{eq:cubelaw_exact}
  y = \frac{x^k}{x^k + (1-x)^k}
  \qquad\hbox{ and in particular, }\qquad
  \frac{dy}{dx}\Bigg{\vert}_{x=\frac{1}{2}} = k .
\end{equation}
It is principally this last result that has an empirical basis: for a party winning around 50\% of all the votes, each additional 1 percent of vote-share translates into an additional $k$ percent of the parliamentary seats.

\begin{figure*}
\caption{Seats versus votes when $p=0.5$, initial conditions 1:1}
\label{fig:cubelaw_exact}
\includegraphics[width=14cm]{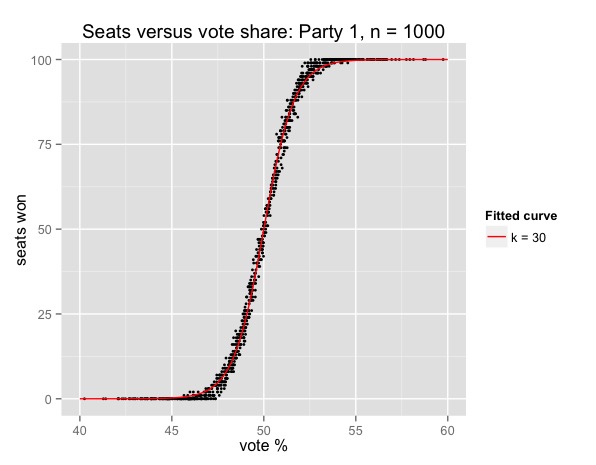}
\end{figure*}

The two-party version of our model also exhibits behaviour of this kind, with the value of $k$ depending on the inter-district imitation parameter $p$. Figure~\ref{fig:cubelaw_exact} shows our results when initially there is 1 voter of each of 2 parties in each district, and $p=0.5$. The fitted (red) line corresponds to the formula \eqref{eq:cubelaw_exact} with $k=30$. The full curve is visible only in situations like this one where strong self-reinforcement ({\it i.e.} large $p$) makes extreme (``landslide'') election outcomes possible.

In general, $k$ increases with $p$: imitation between districts increases the advantage in seats to a party of gaining extra vote share overall, as can be seen in Figure \ref{fig:seatvote_init11}.
This seems intuitively reasonable: even ``useless'' extra voters in districts that their chosen party has no real chance of winning (or no real chance of losing) may still do the party some good if they can influence other voters elsewhere on the electoral map.

Table~\ref{t:slope} shows the best fit for $k$ obtained by ordinary linear regression, for various initial conditions and values of $p$.
The value $k=3$ corresponds then to a value of $p$ slightly more than 0.1.
Note that each additional increase of one unit in popular vote share gives an increase in seat share that is more than one unit, even when the districts are independent ($p=0$).

\begin{table*}
\caption{Best fit slope at $(50,50)$}
\label{t:slope}
\begin{tabular}{ccccc }
\hline
$p$  &  initial conditions 1:1 & initial conditions 2:2 & initial conditions 2:1/1:2 & initial conditions 3:1/1:3 \\
\hline
    0.0 &1.48 &1.67 &1.52& 1.25\\
    0.1 &2.59 & 2.91 & 2.58 & 2.10 \\
    0.2 &4.79 &5.25 &4.82 & 4.29\\
    \hline
\end{tabular}
\end{table*}

For the simplest independent-district cases, it is possible to do some direct calculations of the model's seat-vote relationship. Let $X_1,\ldots,X_N$ be independent, identically distributed random variables representing a party's vote shares in each of $N$ identically-sized districts. Assume that the common distribution of the $X_i$ is the uniform distribution; this corresponds to the large-$n$ limit of the initial model depicted in Figure \ref{fig:votes_init11}. Let $\bar X_N=\frac{1}{N}\sum_{i=1}^N X_i$, the popular vote share. Then the theoretical expected seat-share $y$ corresponding to a popular-vote share $x$ is
\begin{equation}
\label{eq:exactseatvote}
y = P\left(X_1 > \frac{1}{2} \;\Bigg|\; \bar X_N = x\right)
\end{equation}
(since the expected seat share is the same as the probability of winning the first (or any given) district). For $N=2$, \eqref{eq:exactseatvote} yields the exact result
\begin{equation}
y = \begin{cases}    0 &, \mbox{if } x\leq\frac{1}{4}\\
      1 - \frac{1}{4x} &, \mbox{if } \frac{1}{4}\leq x\leq\frac{1}{2}\\
      \frac{1}{4(1-x)} &, \mbox{if } \frac{1}{2}\leq x\leq\frac{3}{4}\\
                     1 &, \mbox{if } x\geq\frac{3}{4}.\\
    \end{cases}
\end{equation}
From this, and a similar but lengthier calculation for $N=3$, we obtain
\begin{equation}
\frac{dy}{dx}\Bigg{\vert}_{x=\frac{1}{2}} =
   \begin{cases}  1 &, \mbox{if } N=2\\
                  2 &, \mbox{if } N=3.\\
   \end{cases}
\end{equation}
Our simulations suggest that these are extreme cases, with the central slope being close to $1.5$ for $N\geq4$.

\subsection{Inter-election swing}
\label{ss:swing}

In many situations, repeated opinion polling at the level of districts is infeasible. Instead, preference changes at district level are often imputed from the overall national \emph{swing} using a model. The most common are the \emph{uniform swing} model and the \emph{proportional swing} model. These models are used, for example, for election prediction --- knowing the exact district-level information from a base election and an estimate of the national swing, we infer the district level information for the next election and hence the election result. 

We can test these models against our simulated data as follows. We first run the urn process with 2 parties and 100 districts, starting with 1 voter of each party in each district, until we have 1000000 voters. We then rescale down to 600 voters, keeping the party percentages the same in each district (up to rounding error) and re-run the process. This typically results in a small difference in each district and overall, which gives the local and overall swing. We did this 1000 times and recorded the results from District 1 each time, yielding 1000 data points.

\begin{figure*}
\caption{Local minus national swing versus original vote share in District 1: $p=0$ (L) and $p=0.1$ (R)} 
\label{fig:swing}
\includegraphics[width=7cm]{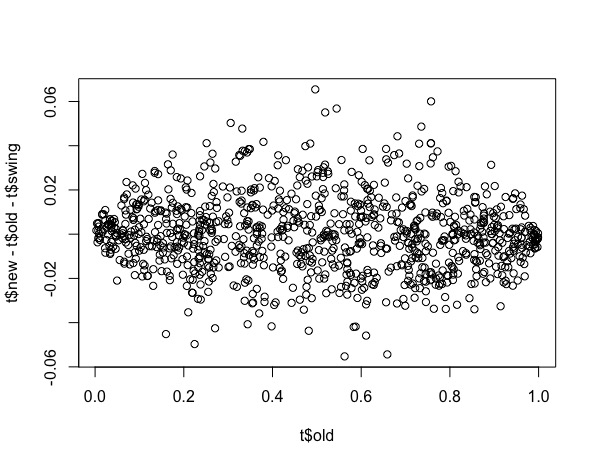}
\includegraphics[width=7cm]{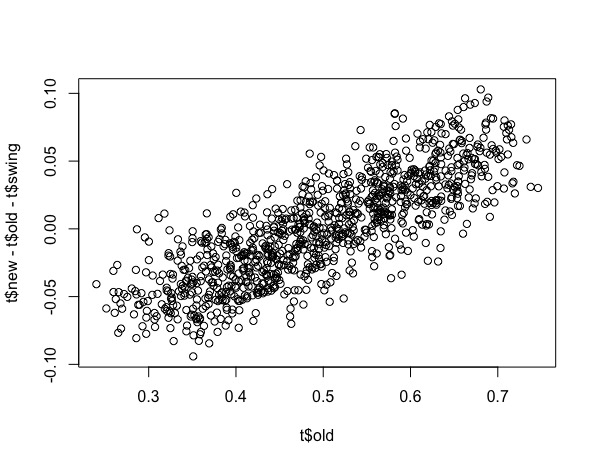}
\end{figure*}

To analyse the data, we plotted the difference between local and national swing against the original vote share in the district. Results are shown in Figure~\ref{fig:swing}. These show clearly that uniform swing fits better for $p=0$ but proportional swing fits better for $p=0.1$. Under uniform swing we expect a horizontal line, while under proportional swing a line with positive slope, moving from negative to positive values.

\section{Conclusion}
\label{s:conc}

Our results above demonstrate the flexibility and power of the model --- with relatively few natural parameters we are able to reproduce a wide range of observed behaviour. In particular with 2 parties, symmetric initial conditions, and $p=0.1$ we can obtain rather realistic-looking data for FPP elections (compare with empirical data shown in \cite{GT79}, for example). This allows for a variety of applications to the analysis of electoral systems using plurality ballots. This includes systems based on multi-member districts in addition to the ``First Past the Post" setup considered in the present paper. For example, we plan a theoretical test of the tradeoff between decisiveness and proportionality of such systems. 

Another use of simulated data generated via the method of this paper is to the analysis of the vote-seat mapping when multi-member districts are involved. Taagepera \cite{Taag1986} generalized the basic cube law formula to the multi-member district case, and one could carry out a best fit analysis similar to that in Section~\ref{ss:cube} for that situation.

One possible extension to the model is to remove the symmetry in the way imitation occurs. For example, some districts may be more influential than others. Or, the sphere of influence of a district may extend only locally -- to other districts geographically or demographically close to it -- rather than to all other districts as assumed in the present paper. In general, an influence structure could be modelled by a graph whose nodes are districts, with imitation probabilities on the (directed) edges. Something like this is probably needed to generate a model of the ``vanishing marginals'' phenomenon observed most strongly in the United States, in which the districts self-organize (or are exogenously organized) into blocs favouring one party or the other, with very few competitive marginal districts. For models including third parties, another possibility is a regionally-based party with supporters who are less likely than most to imitate voters in other districts, or from other parties.

Another possible extension, which we also leave for later, is to consider more complicated preferences. For example, the model works with an arbitrary number of ``colours", each of which represented a party in this article. Instead, each colour could be a linear preference order. In that case, however, it is less reasonable to assume that each change from a given colour to another is equally likely, and a model that incorporates information about the structure of the preferences may be more useful. For example, a transposition of two adjacent elements of the preference order may be more likely than reversing the order entirely.

Finally, we note that a variant of our model in which the total voting population of each district remains fixed, and voters periodically change their state (\emph{i.e.} opinion) by imitating others as in the present paper, is essentially a version of the well-known \emph{voter model} (\cite{L85}). The standard voter model is defined on the integer lattice and emphasizes very local interactions between voters (each voter is influenced only by a handful of immediate neighbours). The possibility of a more general version defined on a graph, with a voter's range of immediate influence extending throughout a large ``district'' and even beyond, has not, to the authors' knowledge, been explored.

\bibliographystyle{plain}
\bibliography{urns}

\begin{thebibliography}{10}

\bibitem{BeLe1994}
Sven Berg and Dominique Lepelley.
\newblock On probability models in voting theory.
\newblock {\em Statistica Neerlandica}, 48(2):133--146, 1994.

\bibitem{BM73}
D.~Blackwell and J.~MacQueen.
\newblock Ferguson distributions via {P}\'{o}lya urn schemes.
\newblock {\em Annals of Statistics}, 1(2):353--355, 1973.

\bibitem{EP23}
F.~Eggenberger and G.~P\'olya.
\newblock {\"U}ber die statistik vorketteter vorg{\"a}nge.
\newblock {\em Zeitschrift f{\"u}r Angewandte Mathematik und Mechanik},
  3(4):279--289, 1923.

\bibitem{GT79}
G.~Gudgin and P.J. Taylor.
\newblock {\em Seats, votes, and the spatial organisation of elections}.
\newblock Pion, London, 1979.

\bibitem{KS50}
M.G. Kendall and A.~Stuart.
\newblock The law of the cubic proportion in election results.
\newblock {\em British Journal of Sociology}, 1:183--196, 1950.

\bibitem{L85}
T.~M. Liggett.
\newblock {\em Interacting particle systems}.
\newblock Springer, New York, 1985.

\bibitem{RGMT2006}
Michel Regenwetter, Bernard Grofman, Anthony Marley, and Ilia Tsetlin.
\newblock Behavioral social choice.
\newblock {\em Cambridge University Press}, 13:58--68, 2006.

\bibitem{Taag1973}
Rein Taagepera.
\newblock Seats and votes: A generalization of the cube law of elections.
\newblock {\em Social Science Research}, 2(3):257--275, 1973.

\bibitem{Taag1986}
Rein Taagepera.
\newblock Reformulating the cube law for proportional representation elections.
\newblock {\em American Political Science Review}, 80(02):489--504, 1986.

\bibitem{T73}
E.~R. Tufte.
\newblock The relationship between seats and votes in two-party systems.
\newblock {\em The American Political Science Review}, 67(2):540--554, 1973.

\end{thebibliography}

\end{document}